\documentclass[%
reprint,
superscriptaddress,
twocolumn,
showpacs,preprintnumbers,showkeys,
amsmath,amssymb,
aps,pre,
]{revtex4-1}

\usepackage{graphicx}
\usepackage{dcolumn}
\usepackage{bm}
\usepackage[latin1]{inputenc}
\usepackage{color}
\newcommand{\rn}[1]{\romannumeral #1}
\newcommand{\RN}[1]{\expandafter\@slowromancap\romannumeral #1@}

\begin{document}

\preprint{APS/123-QED}

\title{Approximate solution for frequency synchronisation in a finite-size Kuramoto model}

\author{Chengwei Wang}
\affiliation{Institute for Complex Systems and Mathematical Biology, University of Aberdeen, King's College, AB24 3UE Aberdeen, United Kingdom.}
\author{Nicol{\'a}s Rubido}
\affiliation{Instituto de F{\'i}sica, Facultad de Ciencias, Universidad de la Rep{\'u}blica, Igu{\'a} 4225, 11400 Montevideo, Uruguay.}
\affiliation{Institute for Complex Systems and Mathematical Biology, University of Aberdeen, King's College, AB24 3UE Aberdeen, United Kingdom.}
\author{Celso Grebogi}
\affiliation{Institute for Complex Systems and Mathematical Biology, University of Aberdeen, King's College, AB24 3UE Aberdeen, United Kingdom.}
\author{Murilo S. Baptista}
\affiliation{Institute for Complex Systems and Mathematical Biology, University of Aberdeen, King's College, AB24 3UE Aberdeen, United Kingdom.}
%
%
%
%

\date{\today}

\begin{abstract}
Scientists have been considering the Kuramoto model to understand the mechanism behind the appearance of collective behaviour, such as frequency synchronisation (FS) as a paradigm, in real-world networks with a finite number of oscillators.
A major current challenge is to obtain an analytical solution for the phase-angles.
Here, we provide an approximate analytical solution for this problem by deriving a master solution for the finite-size Kuramoto model,
without imposing any restriction on the distribution of the natural frequencies of the oscillators. 
The master solution embodies all particular solutions of the finite-size Kuramoto model for any frequency distribution and coupling strength larger than the critical one. 
Furthermore, we present a criterion to determine the stability of the FS solution.
This allows one to analytically infer the relationship between the physical parameters and the stable behaviour of networks.
\end{abstract}
\maketitle

Networks of coupled oscillators provide a pragmatic model to describe basic behaviour of natural and technological systems, such as biological networks \cite{bio.network.1, bio.network.2}, social networks \cite{social.network.1}, computer networks \cite{computer.network.1, computer.network.2} and  power grids \cite{me.models.morden.powergrid, murilo.kuramoto.powergrid}.
A significant phenomenon emerging from coupled oscillators is the synchronisation of the oscillators' rhythms \cite{Mark2008, Dirk2004, phase.synchronisation.1, phase.synchronisation.2}. 
Since 1958, when Norbert Wiener \cite{wiener1958} suggested the presence of collective behaviour of neurons in the brain, 
finding the mechanism and the conditions for the appearance of synchronisation in complex networks has attracted attention of many scientists.
Kuramoto \cite{kuramoto1975, kuramoto1984, kuramoto1987} proposed a mathematically tractable model by considering a network of phase oscillators coupled with an all-to-all topology through a sine function, with each oscillator possessing a constant natural frequency.
Recently, scientists devoted great effort to understand the collective behaviour and synchronisation of the Kuramoto model and its connections to natural systems \cite{kuramoto.small.world, kuramoto.scale.free, politi2015, kuramoto.ring, luccioli2010, Ali2004, kuramoto.chaos}.

However, most of the existing analytical results have considered networks with an infinite number of oscillators and have imposed restrictions on the distributions of natural frequencies of oscillators.
To the best of our knowledge, no analytical work has been proposed  that allows one to determine the synchronous solution of the phase-angles for the finite-size Kuramoto model without restriction on natural frequencies. 
Although the conditions for the set of natural frequencies that provides stable synchronisation is obtained in \cite{Chaos.fully}, the solution for the phase-angles is not given.
But, solving the phase-angles is of fundamental significance in real-world systems. 
For example, in power grids \cite{me.models.morden.powergrid, murilo.kuramoto.powergrid}, which can be described by a Kuramoto-like model \cite{kuramoto.like.power.grid}, one is not only interested in understanding under which conditions the frequency synchronisation (FS) among oscillators emerges, but also in knowing the phase-angles after synchronisation.
The phase-angles are important variables for monitoring generators and developing control strategies for the power grid. 
Similarly, in the research of Josephson junctions \cite{Josephon.junctions1, Josephon.junctions2}, where a Kuramoto-Sakaguchi model is considered, the phase-angles of a synchronous state contain the information of the wave-function phase difference across every Josephson junction.

In this work, we present an analytical method to approximately calculate the phase-angles for the finite-size Kuramoto model when the frequencies are synchronised.
Compared to other works, our method does not require any restriction on the distribution of the natural frequencies of oscillators.
Our solution, shown in Eq.~(\ref{eq:final.solution}), links directly the FS solution and the physical parameters of the network.
Remarkably, the solution is independent of the network size, and only depends on the natural frequencies and the coupling strength.
In addition, we provide an approximate criterion, shown in Eq.~(\ref{criterion.stability.1}), to analytically predict whether a finite number of oscillators are able to emerge into a stable FS, even without knowing the FS solution explicitly.
%

We use $\vec{1}_N~(\vec{0}_N)$ to denote the $N\times 1$ vector with all elements equal to one (zero), $\mathcal{I}_N$ to indicate the index set $\{1,2,\cdots,N\}$, and $\mathbb{R}$ to represent the set of real numbers.
Given a vector $\vec{\alpha}$ with $N$ elements, we use $\overline{\alpha}=\frac{1}{N}\sum_{i=1}^N \alpha_i$ to denote the mean value of $\vec{\alpha}$.

The finite-size Kuramoto, describing the dynamical behaviour of $N$ phase oscillators in an all-to-all network, is given by the equation,
\begin{equation}\label{Origin.Kuramoto}
\dot{\Theta}_i=\Omega_i+\frac{K}{N}\sum_{j=1}^{N}\sin (\Theta_j-\Theta_i),
~~~~i\in \mathcal{I}_N,
\end{equation}
where $N$ is a finite positive integer number, $K$ is the coupling strength, and the $N\times1$ vectors, $\vec{\Omega}=[\Omega_1,~\Omega_2,~\cdots,~\Omega_N]^T$ and $\vec{\Theta}=[\Theta_1,~\Theta_2,~\cdots,~\Theta_N]^T$, denote the natural frequencies and instantaneous phases of the oscillators, respectively.
%
We rewrite Eq.~(\ref{Origin.Kuramoto}) in a rotating frame by letting $\vec{\theta}\equiv\vec{\Theta}-\vec{1}_N \overline{\Omega} t$ and $\vec{\omega}\equiv\vec{\Omega}-\vec{1}_N \overline{\Omega }$, such that $\overline{\omega} =0$. Thus, Eq.~(\ref{Origin.Kuramoto}) becomes
\begin{equation}\label{Kuramoto}
\dot{\theta}_i=\omega_i+\frac{K}{N}\sum_{j=1}^{N}\sin (\theta_j-\theta_i),
~~~~i\in \mathcal{I}_N.
\end{equation}

The order parameter $r\in [0,1]$ is defined by \cite{kuramoto1975, kuramoto1984}
\begin{equation}\label{order}
re^{i\psi}=\frac{1}{N}\sum_{j=1}^{N}e^{i\theta_j},~~~~i\in \mathcal{I}_N,
\end{equation} 
where $r$ and $\psi$ are calculated by equating the real and imaginary parts in Eq.~(\ref{order}), namely,
\begin{equation}\label{r}
r=\sqrt{\left(\frac{1}{N}\sum_{j=1}^N \sin{\theta_j}\right)^2+\left(\frac{1}{N}\sum_{j=1}^N \cos{\theta_j}\right)^2},
\end{equation}
and
\begin{equation}\label{global.angle}
\psi=\arctan\left(\frac{\sum_{j=1}^N \sin{\theta_j}}{\sum_{j=1}^N \cos{\theta_j}}\right).
\end{equation}
We note that $\psi$ is not the average phase in a network with finite oscillators (i.e., $\psi \neq \overline{\theta}$). 
This is different from the situation with infinite number of oscillators \cite{acebron2005, Daniels2005}, where $\psi=\overline{\theta}$ is the global mean field.

Multiplying $e^{-i\theta_i}$ to both sides of Eq.~(\ref{order}) and equating the imaginary part, we have 
\begin{equation}\label{order.tran.1}
r\sin{(\psi-\theta_i)}=\frac{1}{N}\sum_{j=1}^{N}\sin{(\theta_j-\theta_i)},
\end{equation}
which results in
\begin{equation}\label{Kuramoto2}
\dot{\theta}_i=\omega_i+Kr\sin (\psi-\theta_i),~~~~i\in \mathcal{I}_N.
\end{equation}
%
%
The oscillators described by Eq.~(\ref{Kuramoto}) emerge into FS if \cite{Nikhil2005}:
\begin{equation}\label{syn}
\dot{\theta}_i-\dot{\theta}_j = 0 ~\text{as}~ t\rightarrow \infty,~\forall i,j\in \mathcal{I}_N.
\end{equation}
We have, from Eq.~(\ref{Kuramoto}), that $\sum_{i=1}^N\dot{\theta}_i=\sum_{i=1}^N\omega_i=N\overline{\omega}$.
Since $\overline{\omega} =0$, to satisfy Eq.~(\ref{syn}) we require that $\dot{\vec{\theta}}= \vec{1}_N\overline{\omega}=\vec{0}_N$.
Thus, in the rotating framework, the finite-size Kuramoto model in FS can be described by
\begin{equation}
\label{eq.Kuramoto.FS}
\left\lbrace
\begin{split}
&\dot{\theta}_i=\omega_i+\frac{K}{N}\sum_{j=1}^{N}\sin (\theta_j-\theta_i),\\
&\dot{\theta}_i=0,~~~~\forall i\in \mathcal{I}_N.
\end{split}
\right.
\end{equation}
In general, the finite-size Kuramoto model reaches FS if the coupling strength is larger than a critical value, $K_C$ \cite{Ali2004}.
Verwoerd and Mason \cite{Mark2008} provided an algorithm to exactly calculate $K_C$ from
\begin{equation}\label{KC}
K_C=\frac{u}{\frac{1}{N} \sum_{j=1}^N \sqrt{1-\left(\frac{\omega_j}{u}\right)^2}},
\end{equation}
where $u$ is calculated from
$
2\sum_{j=1}^N \sqrt{1- \left( \omega_j/u \right)^2}= \sum_{j=1}^N 1/ \sqrt{1- \left( \omega_j/u \right)^2}.
$
%

The stability of the FS solution of the finite-size Kuramoto model can be studied by the Lyapunov function,
\begin{equation}\label{Lyapunov}
E=\frac{1}{2}\dot{\vec{\theta}}^{\,T} \dot{\vec{\theta}},
\end{equation}
where the time derivative of $E$ along the trajectories of Eq.~({\ref{Kuramoto}}) is
$
\dot{E}=-\frac{K}{N}\sum_{1\leqslant i<j\leqslant N}^N \cos (\theta_i-\theta_j)(\dot{\theta}_i-\dot{\theta}_j)^2.
$
A sufficient condition for the stability of the FS solution is that $\dot{E}<0$, implying $\cos (\theta_i-\theta_j)>0 ,~\forall i,j\in \mathcal{I}_N$.
This means that $|\theta_i-\theta_j|<\frac{\pi}{2},~\forall i,j\in \mathcal{I}_N$.
We denote the stable critical coupling strength of the finite-size Kuramoto model by $K_S$ defined as
\begin{equation}
\label{eq:KS}
K_S:=\min\{K:K\geqslant K_C,~|\theta_i-\theta_j|<\frac{\pi}{2},~\forall i,j\in \mathcal{I}_N\}.
\end{equation}
Thus, as $t\rightarrow \infty$, the oscillators are attracted to the stable FS if $K\geqslant K_S$.
We define $K_S\geqslant K_C$, since $K=K_C$ only ensures the existence of the FS solution \cite{Mark2008}, but $K\geqslant K_S$ provides a condition for its stability based on the Lyapunov function.
The existence of $K_S$ has been studied in \cite{Nikhil2005}, \cite{Dirk2004} and \cite{Ali2004}.

Given $K=K^*\geqslant K_S$, assume $\vec{\theta}^s$ is a stable solution of the finite-size Kuramoto model in FS.
Then, $\vec{\theta}^\xi=\vec{\theta}^s+\vec{1}_N\xi,~\forall \xi \in \mathbb{R}$ is also a stable solution corresponding to $K=K^*$, since the phase differences are independent of $\xi$ .
This means that there are an infinite number of stable solutions for the finite-size Kuramoto model in a certain FS state.
We use $\mathbb{S}_{K^*}^\infty$, corresponding to $K=K^*$, to denote the infinite-dimensional stable solution space.
Actually, $\vec{\theta}=\vec{1}_N \xi,~\forall \xi \in \mathbb{R}$, is the homogeneous solution of Eq.~(\ref{eq.Kuramoto.FS}), obtained by setting the non-homogeneous terms to be zero, i.e., $\vec{\omega}=\vec{0}_N$.
$\vec{\theta}^s$ and $\vec{\theta}^\xi$ are particular solutions of the non-homogeneous Eq.~(\ref{eq.Kuramoto.FS}). 
Our goal is to find one of the particular solutions, which can be analytically expressed, such that Eq.~(\ref{eq.Kuramoto.FS}) is analytically solvable.
We define the master solution of Eq.~(\ref{eq.Kuramoto.FS}) as 
\begin{equation}
\label{eq:master solution}
\vec{\phi}^*=\vec{\theta}^\xi-\vec{1}_N\psi^\xi \in \mathbb{S}_{K^*}^\infty,
\end{equation}
where $\psi^\xi$ is calculated from Eq.~(\ref{global.angle}) as $\vec{\theta}=\vec{\theta}^\xi$.
There are three characteristics for which we call $\vec{\phi}^*$ as the master solution: (\rn{1}) $\vec{\phi}^*$ is analytically expressible; (\rn{2}) $\vec{\phi}^*$ is identical $\forall \xi \in \mathbb{R}$, i.e., $\vec{\phi}^*$ is independent of $\xi$; (\rn{3}) $\psi^*\equiv 0$, where $\psi^*$ is calculated from Eq.~(\ref{global.angle}) as $\vec{\theta}=\vec{\phi}^*$.
Next, we will prove the three characteristics of $\vec{\phi}^*$.

Considering Eq.~(\ref{order.tran.1}), Eq.~(\ref{eq.Kuramoto.FS}) can be transformed into
\begin{equation}\label{eq.to.solve}
\omega_i=K^*r^* \sin \phi_i^*,~\forall i\in \mathcal{I}_N,
\end{equation}
where $r^*$ is calculated by multiplying $e^{-i\psi^\xi}$ on both sides of Eq.~(\ref{order}), namely,
\begin{equation}\label{order.tran.2}
r^*=\frac{1}{N}\sum_{j=1}^N\cos (\theta_j^\xi-\psi^{\xi})=\frac{1}{N}\sum_{j=1}^N\cos (\phi_j^*).
\end{equation}
We define $\Upsilon_{\vec\phi^*}=[\phi_{min}^*,\phi_{max}^*]$, where $\phi_{min}^*$ ($\phi_{max}^*$) is the minimum (maximum) $\phi^*$. 
We have two observations about $\Upsilon_{\vec\phi^*}$: firstly, its length $|\Upsilon_{\vec\phi^*}|<\frac{\pi}{2}$ due to $\vec{\phi^*}\in \mathbb{S}^\infty_{K^*}$; 
secondly, $0\in \Upsilon_{\vec\phi^*}$, because $\overline{\omega} =0$ implies $\omega_{min}\leqslant 0$ and $\omega_{max}\geqslant 0$, resulting in $\phi_{min}^*\leqslant 0$ and $\phi_{max}^*\geqslant 0$ from Eq.~(\ref{eq.to.solve}).
These two characteristics of $\Upsilon_{\vec\phi^*}$ restrict $\Upsilon_{\vec\phi^*}\subset [-\frac{\pi}{2},\frac{\pi}{2}]$.
Thus, we get the analytical expression of the master solution ($\vec{\phi}^*$) from Eq.~(\ref{eq.to.solve}),
\begin{equation}
\label{eq:master solution.analytical}
\phi_i^*=\arcsin \frac{\omega_i}{K^*r^*},~\forall i\in \mathcal{I}_N,
\end{equation}
which is independent of $\xi$, 
thus, (\rn{1}) and (\rn{2}) follows.
From Eq.~(\ref{eq.to.solve}) we have $\sum_{i=1}^N \sin \phi^*_i=\frac{1}{K^*r^*}\sum_{i=1}^N \omega_i=0$.
Substituting this into Eq.~(\ref{global.angle}), we have $\psi^*\equiv 0$, which proves (\rn{3}).
%
%
%
Our goal is to find the master solution ($\vec\phi^*$) as $K=K^*\geqslant K_S$ for the finite-size Kuramoto model.
We know $\Upsilon_{\vec\phi^*}\subset [-\frac{\pi}{2},\frac{\pi}{2}]$,
implying $\cos \phi_j\geqslant 0,~\forall j\in \mathcal{I}_N$.
Thus, the order parameter in Eq.~(\ref{order.tran.2}) is calculated as $r^*=\frac{1}{N} \sum_{j=1}^N \sqrt{1-\sin (\phi_j^*)^2}$. Considering $\sin \phi_j^*=\omega_j/(K^*r^*)$ obtained from Eq.~(\ref{eq.to.solve}), $r^*$  can be expressed by a transcendental equation, namely,
\begin{equation}
\label{eq:order.exact}
r^*=\frac{1}{N} \sum_{j=1}^N \sqrt{1-\left(\frac{\omega_j}{K^*r^*} \right)^2}.
\end{equation} 
In order to obtain an approximate solution for $r^*$, we construct a new model for the original system.
We relabel $\vec\omega$ such that $\omega_1\leqslant\omega_2\leqslant\cdots\leqslant\omega_N$,
and split $\omega$ into two groups.
One group is $\omega'=[\omega_1,\omega_2,\cdots,\omega_{N'}]^T$, where $N'=\frac{N}{2}~(N'=\frac{N-1}{2})$ if $N$ is even (odd),  
and the other group is $\omega''=[\omega_{N'+1},\omega_{N'+2},\cdots,\omega_N]^T$.  
We note that, since $\sum_{j=1}^{N'}\omega_j+\sum_{j=N'+1}^N \omega_j=\sum_{j=1}^N \omega_j=0$,
$\overline{\omega'} =-\\overline{\omega''}\leqslant 0$ when $N$ is even, and $\overline{\omega'} \approx -\overline{\omega''} \leqslant 0$ when $N$ is odd.
For simplicity, we indistinctly denote $\overline{\omega'} \approx -\overline{\omega''}$ for both cases.

When all of the oscillators emerge into stable FS, our model treats the whole system as two oscillators in stable FS. 
The natural frequencies of the two oscillators are $\overline{\omega' }$ and $\overline{\omega'' }$.
The two oscillators also follow the original FS Kuramoto model equations, namely,
\begin{equation}\label{eq.+}
\overline{\omega' }=K^*{r^{*}}'\sin \phi',
\end{equation}
\begin{equation}\label{eq.-}
\overline{\omega'' }=K^*{r^{*}}'\sin \phi'',
\end{equation}
where ${r^{*}}'$ is the order parameter for the two oscillators.
The two oscillators are in stable FS, thus we have $\phi',~\phi''\in [-\frac{\pi}{2},\frac{\pi}{2}]$, which is obtained from the analysis of the Lyapunov function [Eq.~(\ref{Lyapunov})] for the two oscillators. Then the order parameter is
\footnotesize{
${r^{*}}'=1/2\left(\cos \phi'+\cos \phi''\right)
=1/2 \sqrt{1-\left[\overline{\omega' }/(K^*r^*)\right]^2} +
1/2 \sqrt{1-\left[\overline{\omega'' }/(K^*r^*)\right]^2}
$
}.
\normalsize{ Further considering $ |\overline{\omega' }| \approx |\overline{\omega'' }|$ we have}
\begin{equation}\label{r*'}
{r^{*}}'\approx \sqrt{1-\left(\frac{\overline{\omega' }}{K^*r^*}\right)^2},
\end{equation}
whose solution is
\begin{equation}\label{r'1}
{r^{*}}'_1\approx \lambda_1= \frac{\sqrt{2}}{2}\sqrt{1+\sqrt{1-\frac{4\overline{\omega' }^2}{K^{*2}}}},~K^*\geqslant 2|\overline{\omega' }|,
\end{equation}
\begin{equation}\label{r'2}
{r^{*}}'_2\approx \lambda_2=\frac{\sqrt{2}}{2}\sqrt{1-\sqrt{1-\frac{4\overline{\omega' }^2}{K^{*2}}}},~K^*\geqslant 2|\overline{\omega' }|,
\end{equation}
where we have $\lambda_1\lambda_2=\frac{-\overline{\omega' }}{K^*}\approx \frac{\overline{\omega'' }}{K^*}$.
Substituting this condition into Eq.~(\ref{eq.+}) and Eq.~(\ref{eq.-}), we get $-\lambda_1\lambda_2={r^{*}}'\sin \phi'$ and $\lambda_1\lambda_2={r^{*}}'\sin \phi''$.
If ${r^{*}}'\approx \lambda_2$, we have $\phi'\approx -\arcsin (\lambda_1)$ and $\phi''\approx \arcsin (\lambda_1)$.
Because $\frac{\sqrt{2}}{2}\leqslant\lambda_1\leqslant 1$, we approximately have $-\frac{\pi}{2}\leqslant \phi'\leqslant -\frac{\pi}{4}$ and $\frac{\pi}{4}\leqslant\phi''\leqslant \frac{\pi}{2}$. 
This means $|\phi'-\phi''|\geqslant \frac{\pi}{2}$, and $\lambda_1$ grows larger as $K$ increases from $K_S$ resulting in a growth of $|\phi'-\phi''|$.
However, $|\phi'-\phi''|\geqslant \frac{\pi}{2}$ implies instability of the FS solution of the two oscillators, which can be understood by the Lyapunov function in Eq.~(\ref{Lyapunov}) for the two oscillators.
This means that ${r^{*}}'\approx \lambda_2$ describes an unstable FS solution. 
On the other hand, ${r^{*}}'\approx \lambda_1$ ensures the stability of the FS solution.
Thus, we let $r^* \approx {r^{*}}'\approx \lambda_1$ be the approximation for the order parameter in Eq.~(\ref{eq:order.exact}),
then the approximation ($\vec\phi^{**}$) for the master solution ($\vec\phi^*$) in Eq.~(\ref{eq:master solution.analytical}) is 
\begin{equation}
\label{eq:master solution.Apro}
\phi_i^{**}= \arcsin \frac{\omega_i}{K^*\lambda_1},~\forall i\in \mathcal{I}_N
\end{equation}
Consequently, the approximate stable solution of the finite-size Kuramoto model in FS is 
\begin{equation}
\label{eq:final.solution}
\theta_i \approx \arcsin \frac{\omega_i}{K^*\lambda_1}+ \xi,~\forall i\in \mathcal{I}_N,~\forall \xi \in \mathbb{R}.
\end{equation}
%
%
%
\begin{figure}[h]
\centering
\includegraphics[width=0.35\linewidth]{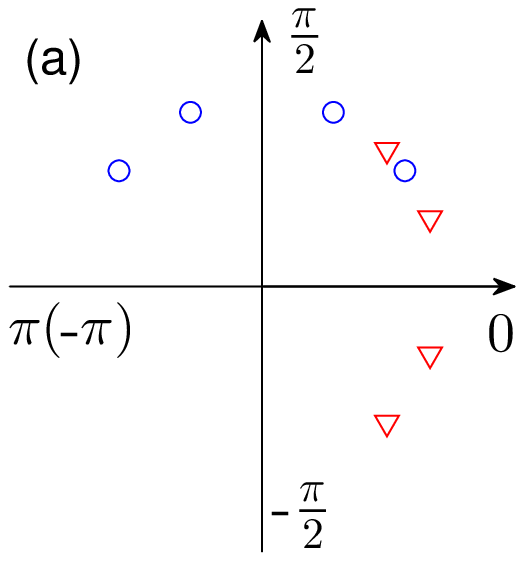}~
\includegraphics[width=0.35\linewidth]{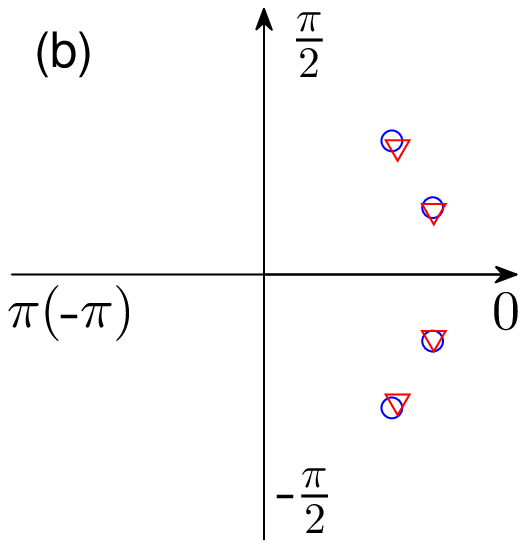}
\includegraphics[width=0.35\linewidth]{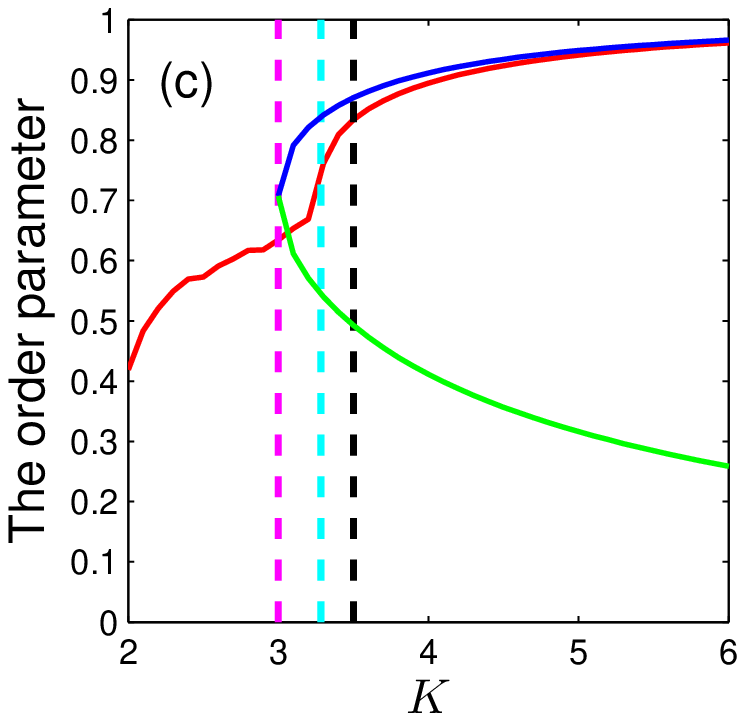}~
\includegraphics[width=0.35\linewidth]{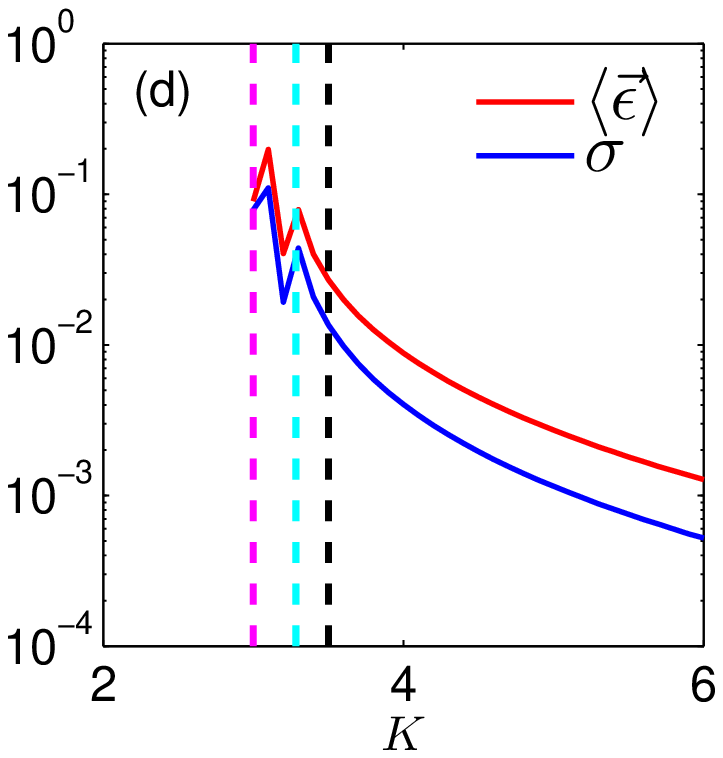}
\caption{[Colour online] Results for a network of $4$ oscillators with the vector of natural frequency given by $\vec\Omega=[-2,-1,1,2]^T$. 
(a) $K<K_S$, the red triangles and blue circles indicate the approximate master solution ($\vec{\phi}^{**}$) in Eq.~(\ref{eq:master solution.Apro}) and the numerical one ($\vec{\phi}^*$) in Eq.~(\ref{eq:master solution.analytical}), respectively. 
(b) $K=K_S$, $\vec{\phi}^{**}$ indicated by red triangles is close to $\vec{\phi}^*$ indicated by blue circles. 
(c) The change of the order parameter and its approximation with respect to $K$. The red solid line indicates the numerical result of the order parameter in Eq.~(\ref{eq:order.exact}), (an average value of results from 2000 simulations with different initial phases).  
The blue solid line and green solid line indicate the change of $\lambda_1$ in Eq.~(\ref{r'1}) and $\lambda_2$ in Eq.~(\ref{r'2}), respectively, as $K\geqslant 2|\overline{\omega'}|$, where $2|\overline{\omega'}|$ is the lowest bound of $K$ in Eq.~(\ref{r'1}) and Eq.~(\ref{r'2}). 
The magenta (cyan, black) dash line represents $2|\overline{\omega'}|~(K_C,~K_S)$. 
(d) The change of the average of the absolute error (red solid line) and standard deviation  of the absolute error (blue solid line) between $\vec{\phi}^{**}$ and $\vec{\phi}^*$ as a function of $K$ when $K\geqslant 2|\overline{\omega'}|$. 
}
\label{fig.4nodes}
\end{figure}
We use $\vec\epsilon$ with element $\epsilon_i=|\phi^{**}_i-\phi^*_i|,~\forall i \in \mathcal{I}_N$, to denote the absolute error between the approximate master solution [$\vec{\phi}^{**}$ in Eq.~(\ref{eq:master solution.Apro})] and the numerical one [$\vec{\phi}^*$ in Eq.~(\ref{eq:master solution.analytical})],
and $\sigma$ to denote the standard deviation of $\vec{\epsilon}$, defined as $\sigma=\sqrt{\frac{1}{N} \sum_{i=1}^N (\epsilon_i-\overline{\epsilon})^2}$.
To demonstrate the effectiveness of our method for a network with four oscillators, we show numerical results in Fig.~\ref{fig.4nodes}. 
$\epsilon_i$ is large $\forall i\in \mathcal{I}_N$, when $K<K_S$ [$\vec\phi^{**}\neq \vec\phi^*$ in Fig.~\ref{fig.4nodes} (a)], and small when $K=K_S$ [$\vec\phi^{**}\approx \vec\phi^*$ in Fig.~\ref{fig.4nodes} (b)].
The approximate order parameter, $\lambda_1$, is close to the numerical one as $K\geqslant K_S$ [Fig.~\ref{fig.4nodes} (c)].
The average absolute error and standard deviation of the absolute error are small ($\overline{\epsilon}<10^{-1},~\sigma<10^{-1}$) when $K\geqslant K_S$, and decreases rapidly as $K$ is increased [Fig.~\ref{fig.4nodes} (d)].
This means that our method can calculate the master solution almost exactly in an analytical way for this network.

\begin{figure}[h]
\centering
\includegraphics[width=0.35\linewidth]{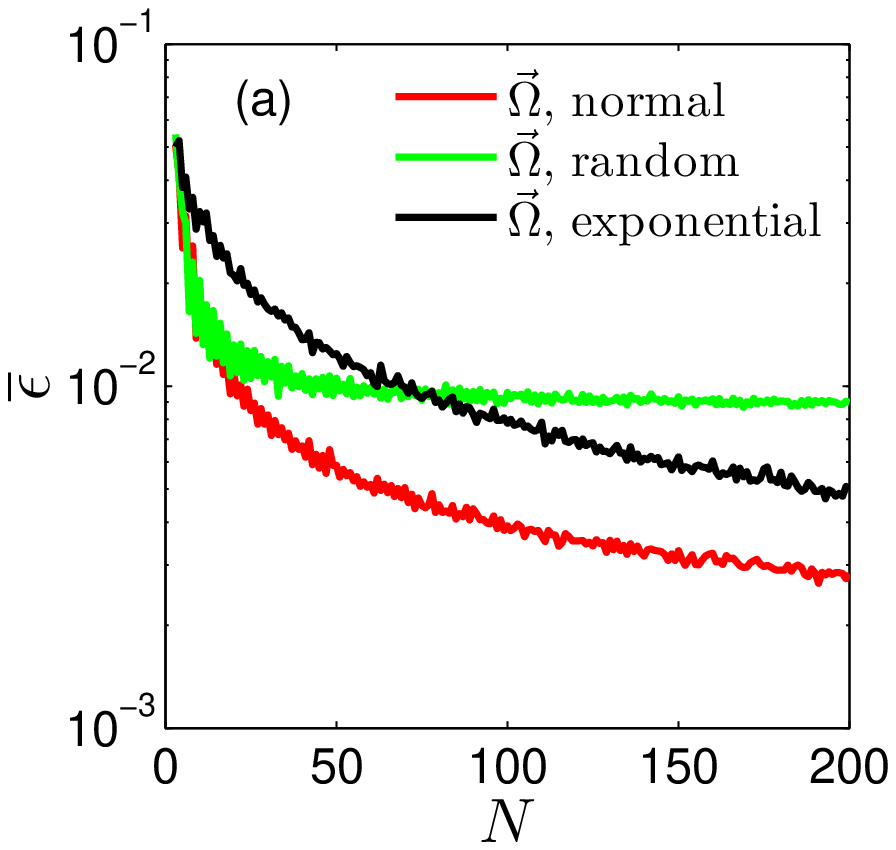}~
\includegraphics[width=0.35\linewidth]{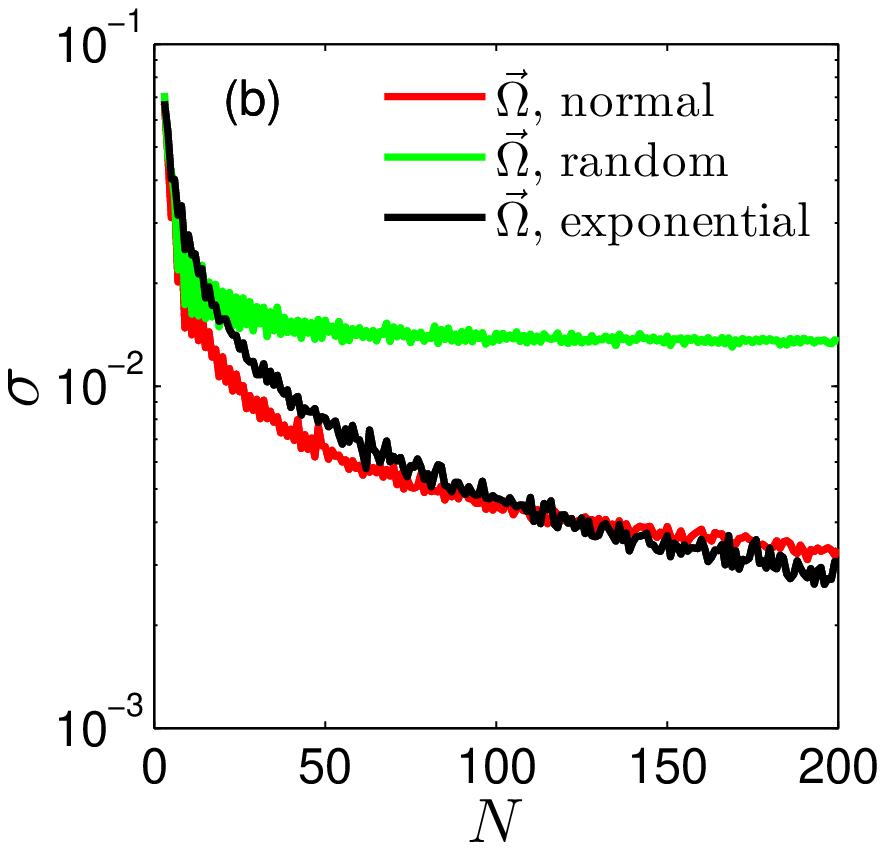}
\caption{[Colour online] Study of the average absolute error $\overline{\epsilon }$ in (a), and the standard deviation $\sigma$ in (b) between the approximate master solution ($\vec{\phi}^{**}$) and the numerical one ($\vec{\phi}^*$) at $K=K_S$. $N$ increases from $3$ to $200$. The red line (green line, black line) corresponds to the normal distribution (random distribution, exponential distribution) of $\vec\Omega$. Results are based on the average value of results from 100 simulations for each distribution.}
\label{fig.error}
\end{figure}

Figure~\ref{fig.error} indicates that our method works well for networks with different  $\vec\Omega$ distributions (random, normal and exponential) and different number of oscillators ($N$ increasing from $3$ to $200$).
We observe that $\overline{\epsilon}$ and $\sigma$ are larger when $N$ is smaller, but they decrease quickly as $N$ increases.
Furthermore, $\overline{\epsilon}$ and $\sigma$ in Fig.~\ref{fig.error} are obtained at $K=K_S$, meaning that they are the largest value obtained for each simulation.
In other words, smaller  $\overline{\epsilon}$ and $\sigma$ can be obtained if we increase $K$ for any given $N$ and any $\vec\Omega$ distribution, since larger $K$ implies smaller absolute error between the approximate order parameter and the numerical one as shown in Fig.~\ref{fig.4nodes} (c), which further implies smaller absolute error between $\vec{\phi}^{**}$ and $\vec{\phi}^*$.
%
%

A sufficient condition to ensure the stability of the FS solution was proposed by \cite{Dirk2004} as
\begin{equation}
\label{eq:stable.ref}
K>K_P=\frac{\sqrt{2}|\omega_i|}{r},~\forall i\in \mathcal{I}_N.
\end{equation}
Taking $\lambda_1$ in Eq.~(\ref{r'1}) as the approximation of $r$, we get a sufficient condition for the stability of the FS solution, namely,
\begin{equation}
\label{criterion.stability.1}
K> K_A= \frac{\sqrt{2}|\omega_i|}{\lambda_1},~\forall i\in \mathcal{I}_N.
\end{equation}
Equation~(\ref{criterion.stability.1}) is useful to approximately forecast whether the system is able to get into a stable FS state in an analytical way, without solving the differential equations.
In other words, it reveals the relationship between the physical parameters (average natural frequency and coupling strength) and the stable behaviour of oscillator networks described by the finite-size Kuramoto model explicitly. 
\begin{figure}[h]
\centering
\includegraphics[width=0.3\linewidth]{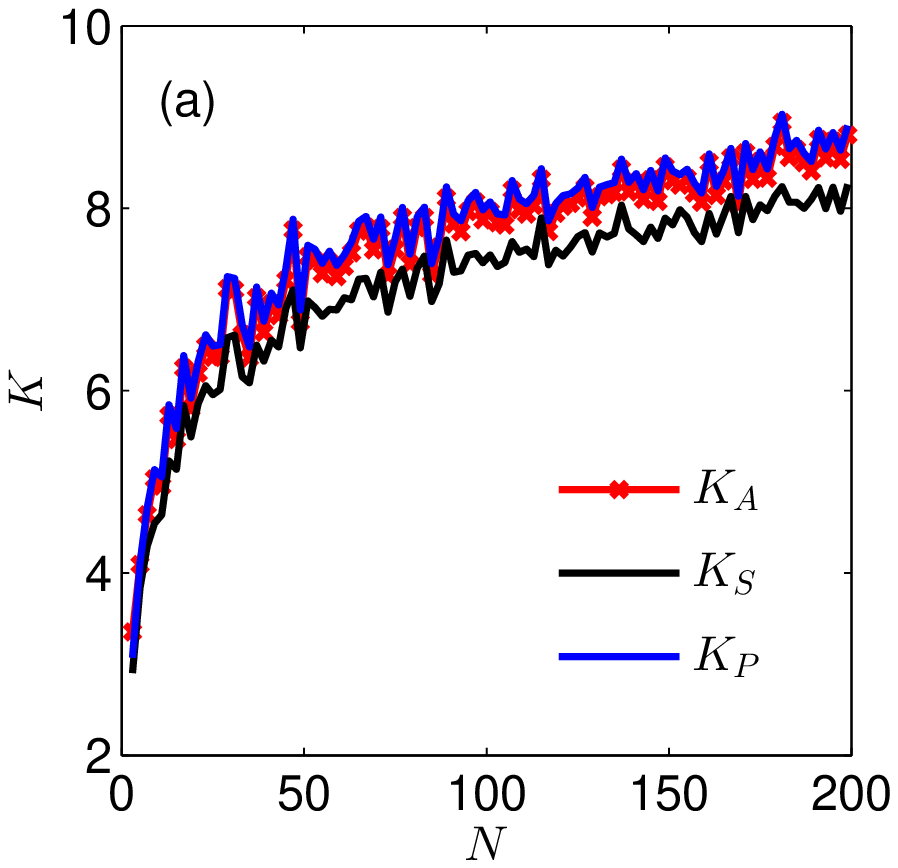}
\includegraphics[width=0.3\linewidth]{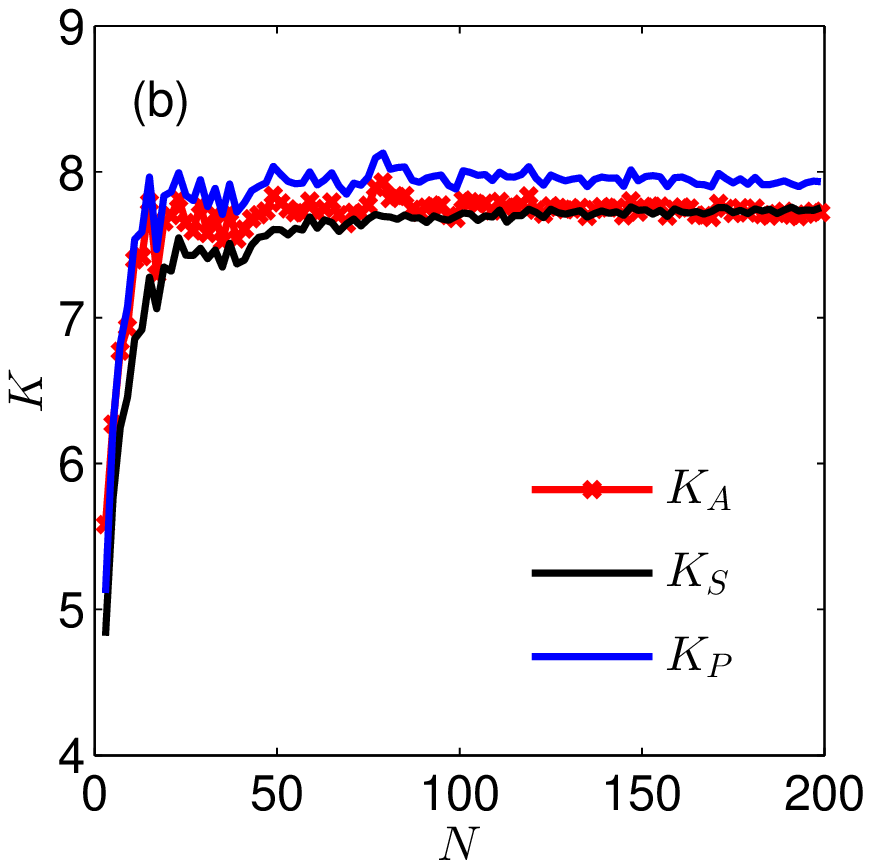}
\includegraphics[width=0.3\linewidth]{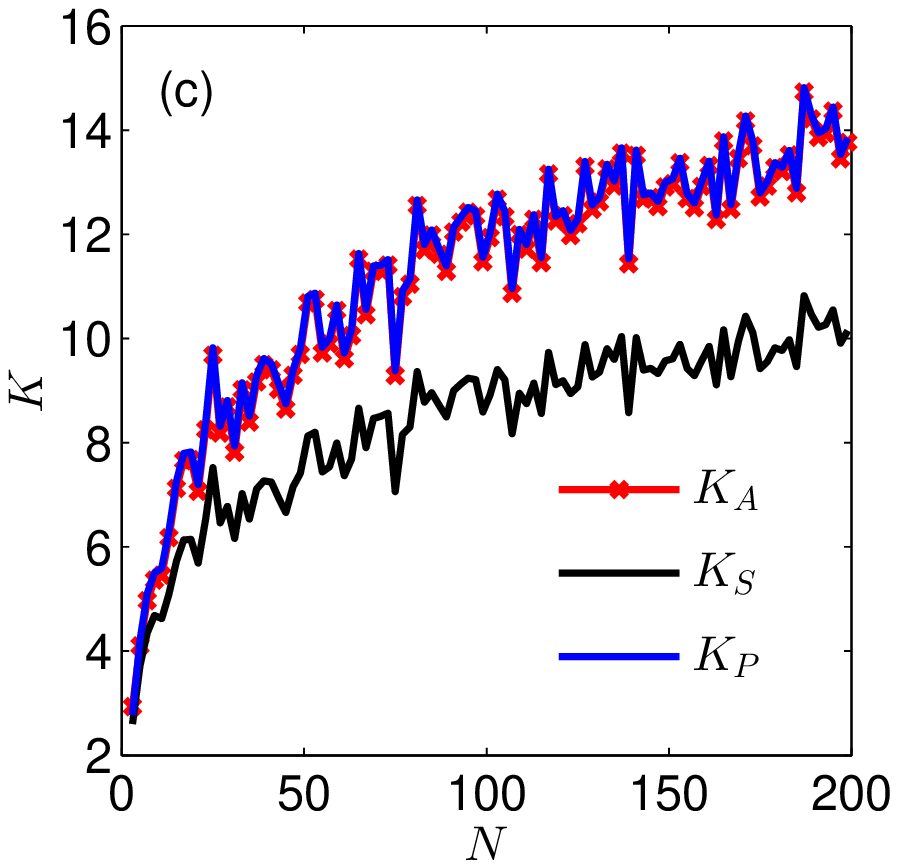}
\caption{[Colour online] The stable critical coupling strength of the finite-size Kuramoto model. $N$ increases from $3$ to $200$. $K_S$ is calculated by Eq.~(\ref{eq:KS}), $K_P$ by Eq.~(\ref{eq:stable.ref}), $K_A$ by Eq.~(\ref{criterion.stability.1}). $(a)$ $\vec\Omega$ follows a normal distribution. $(b)$ $\vec\Omega$ follows a random distribution. $(c)$ $\vec\Omega$ follows an exponential distribution. The figures are drawn based on average results of 30 simulations for each distribution.}
\label{fig.stability}
\end{figure}

Figure~\ref{fig.stability} shows the effectiveness of our condition to gauge the stability for the FS solution of the finite-size Kuramoto model,
considering different $\vec{\Omega}$ distributions and different number of oscillators. 
If $\vec\Omega$ obeys a normal or exponential distribution [Figs.~\ref{fig.stability} (a) and \ref{fig.stability} (c), respectively], $K_A$ in Eq.~(\ref{criterion.stability.1}) coincides remarkably with $K_P$ in Eq.~(\ref{eq:stable.ref}), and $K_A>K_S$.
If $\vec\Omega$ obeys a random distribution [Fig.~\ref{fig.stability} (b)], $K_A$ in Eq.~(\ref{criterion.stability.1}) is close to $K_S$ in Eq.~(\ref{eq:KS}).
This means Eq.~(\ref{criterion.stability.1}) is effective to approximately determine the stability for the FS solution of the finite-size Kuramoto model.

%
In this paper, we studied the finite-size Kuramoto model [Eq.~(\ref{Kuramoto})], including its analytical solution for frequency synchronisation (FS) and its stable behaviour. 
Our approximate FS solution takes a simple form [Eq.~(\ref{eq:final.solution})], 
which, surprisingly, is independent of the network size.
These significant results is a consequence of a mathematical insight expressed in Eq.~(\ref{eq:master solution.analytical}) and a physical insight in the model leading to Eq.~(\ref{eq.+}) and Eq.~(\ref{eq.-}).
Among an infinite number of FS solutions, we have understood that there is a particular one, the master solution in Eq.~(\ref{eq:master solution.analytical}), which allows one to calculate all of the others.
We have also understood that the FS in the finite-size Kuramoto model is approximately characterised by 2 clusters of oscillators that can be effectively described as 2 coupled oscillators in Eq.~(\ref{eq.+}) and Eq.~(\ref{eq.-}).
Furthermore, we developed a condition to approximately predict the stability for the FS solution of the finite-size Kuramoto model in an analytical way. 
This condition allows one to easily infer the relationship between the physical parameters and the stable behaviour of networks.
\bibliography{mybibphy}{}
\end{document}